\documentclass[conference,letterpaper]{IEEEtran}
\IEEEoverridecommandlockouts
\usepackage[utf8]{inputenc} 
\usepackage[T1]{fontenc}
\usepackage{url}
\usepackage{ifthen}
\usepackage{graphicx}
\usepackage{subfigure}
\usepackage[cmex10]{amsmath} 
\usepackage{amssymb,mathtools}
\usepackage{color,soul}
\usepackage{hyperref}  
\hypersetup{colorlinks,
            linkcolor=blue,
            citecolor=blue,
            urlcolor=black,
            anchorcolor=blue,
            filecolor=blue,
            linktocpage,
            plainpages=false,
            breaklinks=true}
\usepackage[style=ieee,maxnames=2, minnames=1, maxbibnames=99,citestyle=numeric-comp]{biblatex}

\usepackage{enumitem}

\newtheorem{theorem}{Theorem}      
\newtheorem{lemma}{Lemma}          
 
\newtheorem{corollary}{Corollary}  
\newtheorem{definition}{Definition} 
\newtheorem{remark}{Remark}        

\newcounter{relctr} 
\everydisplay\expandafter{\the\everydisplay\setcounter{relctr}{0}} 

\AtBeginDocument{} 

\newcommand{\bR}{\ensuremath{\mathbb{R}}}

\newcommand{\cD}{\ensuremath{\mathcal{D}}}

\newcommand{\cI}{\ensuremath{\mathcal{I}}}

\newcommand{\cQ}{\ensuremath{\mathcal{Q}}}

\newcommand{\cX}{\ensuremath{\mathcal{X}}}
\newcommand{\cY}{\ensuremath{\mathcal{Y}}}

\newcommand{\dinf}{\ensuremath{D_\infty}}

\allowdisplaybreaks

\begin{document}
\title{Context-aware Privacy Bounds for Linear Queries} 


\author{%
  \IEEEauthorblockN{Heng Zhao\IEEEauthorrefmark{1},
  Sara Saeidian\IEEEauthorrefmark{1}\IEEEauthorrefmark{2},
  Tobias J. Oechtering\IEEEauthorrefmark{1}}
  \IEEEauthorblockA{\IEEEauthorrefmark{1}KTH Royal Institute of Technology, 100 44 Stockholm, Sweden, hengzhao2026@outlook.com, \{saeidian, oech\}@kth.se}
  \IEEEauthorblockA{\IEEEauthorrefmark{2}Inria Saclay, 91120 Palaiseau, France}
\thanks{This work was supported by the Swedish Research Council (VR) under grants 2023-04787 and 2024-06615.}
}

\maketitle


\begin{abstract}
Linear queries, as the basis of broad analysis tasks, are often released through privacy mechanisms based on differential privacy (DP), the most popular framework for privacy protection.
However, DP adopts a context-free definition that operates independently of the data-generating distribution. 
In this paper, we revisit the privacy analysis of the Laplace mechanism through the lens of pointwise maximal leakage (PML).
We demonstrate that the distribution-agnostic definition of the DP framework often mandates excessive noise.
To address this, we incorporate an assumption about the prior distribution by lower-bounding the probability of any single record belonging to any specific class.
With this assumption, we derive a tight, context-aware leakage bound for general linear queries, and prove that our derived bound is strictly tighter than the standard DP guarantee and converges to the DP guarantee as this probability lower bound approaches zero.
Numerical evaluations demonstrate that by exploiting this prior knowledge, the required noise scale can be reduced while maintaining privacy guarantees.
\end{abstract}

\section{Introduction}

Many modern applications leverage mathematically rigorous frameworks to safeguard user privacy nowadays.
Among these, \emph{differential privacy} (DP)~\cite{DworkCalibratingNoiseSensitivity,DworkFoundations} has emerged as one of the most widely adopted approaches. DP protects privacy by masking the contribution of individual records through the addition of carefully calibrated noise to a mechanism’s output. A fundamental characteristic of DP is its \emph{context-free} nature since its definition does not explicitly depend on the data-generating distribution. This property has prompted two ongoing debates. 
First, prior work has argued that DP may not be well-suited for correlated datasets~\cite{kiferNoFreeLunch2011,Kifer2012framework,apf24}, suggesting that although DP is expressed in a context-free form, it implicitly assumes independence among data records. Second, and central to this work, is the question of utility and the amount of noise required to achieve satisfactory privacy guarantees. Specifically, can a context-aware framework leverage assumptions about the data-generating distribution to better calibrate the noise and thereby achieve higher utility?

\emph{Pointwise maximal leakage} (PML)~\cite{SaeidianPML} provides precisely the operational framework needed to address this question. 
Unlike DP, which focuses on the similarity of a mechanism's outputs, PML is grounded in operational threat models, and quantifies the maximum information gain an adversary can achieve about the secret data relative to their prior belief~\cite{SaeidianRethinking}. 
More precisely, this gain function view of PML can be used to explicitly model a large variety of adversarial goals, e.g., membership inferences or attribute disclosure attacks~\cite{SaeidianPML}.
Importantly, \textcite{SaeidianRethinking} also established a formal equivalence between DP and PML: Theorem 4.2 in~\cite{SaeidianRethinking} proves that satisfying pure $\varepsilon$-DP is equivalent to bounding the PML of every individual record across all possible product distributions. 
This result positions PML not merely as a distinct metric, but as a generalized framework that recovers standard DP as a special instance.

In this paper, we leverage this connection between DP and PML to revisit the privacy guarantees of the standard Laplace mechanism for the broad class of \emph{linear queries}. 
Linear queries are a class of queries that involve computing weighted sums of counts of data records.
They serve as fundamental building blocks for data analytics, spanning from simple aggregations like counting queries, range queries~\cite{Li2012AnAdaptiveMechanism} and sliding window sums, to complex decompositions such as wavelets~\cite{li2015matrix}.
We incorporate an assumption about the prior distribution by lower-bounding the probability of any single record belonging to any specific class, which ensures no data class is arbitrarily rare.
With this assumption, we derive a tight, context-aware leakage bound for linear queries released by the Laplace mechanism.
Our analysis quantifies the "conservatism" of standard DP: we show that when we can restrict the prior distribution, the same level of noise offers stronger privacy protection than the DP parameter suggests.
This work generalizes recent findings regarding histogram publication~\cite{Saeidian2025Histogram}. 
It focused on the special case where the raw counts of disjoint data categories are requested. 
It demonstrated that, privacy guarantees are stronger when the probability mass of each class is bounded away from zero. 
By extending this context-aware analysis to general linear queries, our work is more general and covers histogram publication as a special case.

In many data analysis scenarios, the data curator needs to answer a specific collection of linear queries, collectively referred to as the query workload. 
Existing literature on linear queries has largely focused on mechanism design to minimize error under fixed privacy constraints for such workloads.
Workload-aware strategies like the matrix mechanism~\cite{li2015matrix} exploit the post-processing property of differential privacy.
Rather than answering the target workload directly, they answer an optimized set of queries and linearly combine the noisy results to reconstruct the target answers with minimal error.
High-dimensional matrix mechanism (HDMM)~\cite{HDMM} generalizes the matrix mechanism to high-dimensional setups.
In a separate line of work, the data-aware/workload-aware mechanism (DAWA)~\cite{LiDAWA} uses a portion of the privacy budget to estimate the data distribution, dynamically partitioning the domain by grouping consecutive classes of records with similar counts into buckets. Then DAWA is able to better calibrate noise for releasing range queries.
In another line of work, iterative methods like multiplicative weights exponential mechanism~\cite{MWEM} and adaptive and iterative mechanism (AIM)~\cite{AIM} iteratively construct a synthetic dataset that mimics the statistical properties of the private data.
In this work, rather than designing new mechanisms, we revisit the fundamental privacy analysis of the standard Laplace mechanism under the PML framework. 
The results of this paper can later be extended to more sophisticated frameworks for releasing linear queries.

\textbf{Contributions and Outline.}
First, we derive a tight PML bound for general linear queries released via the Laplace mechanism and also provide a computationally efficient simplified bound (Theorem~\ref{thm:linear_queries_pml} and Corollary~\ref{crl:simplebound}). 
We prove that when the data generating distribution is assumed to assign a minimum probability mass to every class, our bound offers a strictly tighter privacy guarantee than standard DP analysis. 
We further establish that as this minimum probability assumption is relaxed towards zero, our bound converges exactly to the standard DP budget.
Finally, we empirically validate our theoretical findings across various linear query workloads. 
Numerical evaluations demonstrate that by leveraging knowledge about prior distribution, the required noise scale can be reduced compared to standard DP mechanisms, thereby achieving higher utility without compromising privacy.

\section{Preliminaries}
\subsection{Notation}
We denote random variables by capital letters (e.g., $X$), their realizations by lowercase letters (e.g., $x$), and sets by calligraphic letters (e.g., $\mathcal{X}$). We use $[n]$ to denote the set of integers $\{1, 2, \dots, n\}$ and $\log$ to denote the natural logarithm. 
Throughout the paper, $X$ represents the sensitive data with distribution $P_X$, and $Y$ denotes the output of a privacy mechanism (i.e., a conditional distribution) $P_{Y|X}$ with marginal distribution $P_Y$. For notational convenience, we assume that the alphabet of $X$, $\cX$, has full support. 

\subsection{Pointwise Maximal Leakage}
Pointwise maximal leakage (PML) \cite{SaeidianPML,SaeidianPMLonAlp} is an operationally meaningful privacy measure rooted in \emph{quantitative information flow} \cite{alvim2020science}.
PML is derived by analyzing privacy risks within two robust and general adversarial models: the \emph{randomized function model} (introduced in ~\cite{IssaOperational}) and the \emph{gain function model} (introduced in ~\cite{AlvimGain,AlvimMultiplicative}).

Here, we briefly recall the definition of PML using the gain-function threat model. Consider an adversary whose objective is described by a gain function $g : \mathcal{X} \times \mathcal{W} \to \mathbb{R}_+$, where $\mathcal{W}$ denotes the adversary’s guessing space. The value $g(x,w)$ represents the adversary's reward obtained by guessing $w$ when the true value of the secret is $x$. For a given gain function $g$ and outcome $y \in \mathcal{Y}$ of the mechanism $P_{Y \mid X}$, the information leakage is defined as the ratio between the adversary’s posterior expected gain after observing $y$ and their prior expected gain. PML is then obtained by maximizing this ratio over all possible gain functions. This maximization provides robustness, since it accounts for a broad class of adversarial objectives.

\begin{definition}[Pointwise maximal leakage~{\cite[Def.~3]{SaeidianPMLonAlp}}]
\label{def:PML}
Suppose $X \sim P_X$ and let $Y$ be the random variable induced by the mechanism $P_{Y \mid X}$. The pointwise maximal leakage from $X$ to $y \in \mathcal{Y}$ is defined as
\begin{equation}
\label{eq:g-leakage}
    \ell(X \to y) := \log \; \sup_{g, \mathcal{W}} \; \frac{\sup\limits_{P_{W \mid Y}} \mathbb{E}[g(X, W) \mid Y = y]}{\sup_{w' \in \mathcal{W}} \mathbb{E}[g(X, w')]},
\end{equation}
where $P_{W \mid Y}$ is the conditional distribution of the adversary's guess $W$ given $Y$. The supremum is over all measurable spaces $\mathcal{W}$ and non-negative measurable functions $g$ with $\sup_{w' \in \mathcal{W}} \mathbb{E} [g(X,w')] < \infty$.
\end{definition}

For linear queries, the input space $\mathcal{X}$ is finite and the output space $\mathcal{Y}$ is Euclidean. Under this setting, and as shown in \cite{SaeidianPMLonAlp}, Definition~\ref{def:PML} admits the following simple expressions:
\begin{align}
\begin{split}
\label{eq:pml_simple}
    \ell(X \to y) &= \dinf(P_{X \mid Y=y} \Vert P_X ) = \log \max_{x \in \mathcal{X}} \frac{P_{X \mid Y=y}(x)}{P_X(x)}\\
    &= \log \max_{x \in \mathcal{X}} \frac{f_{Y \mid X=x}(y)}{f_Y(y)},
\end{split}
\end{align}
where $\dinf(\cdot \Vert \cdot)$ denotes the Rényi divergence of order $\infty$~\cite{renyi1961measures,van2014renyi}, $P_{X \mid Y=y}$ is the posterior distribution of $X$ given $y$, and 
\begin{equation*}
    f_{Y \mid X=x} = \frac{d P_{Y \mid X=x}}{d \lambda}, \quad f_Y = \frac{d P_Y}{d \lambda},
\end{equation*}
are densities with respect to the Lebesgue measure $\lambda$. 

\begin{remark}
It follows immediately from \eqref{eq:pml_simple} that PML satisfies the trivial upper bound 
\begin{equation}
\label{eq:trivial}
    \ell(X \to y) \leq \log \frac{1}{\min_{x \in \cX} P_X(x)},
\end{equation}
for all mechanisms $P_{Y \mid X}$ and all $y \in \cY$. 
\end{remark}


\subsection{Differential Privacy and Laplace Mechanism}
Differential privacy (DP) requires that a mechanism should produce nearly indistinguishable outputs on \emph{neighboring} databases, i.e., databases that differ by one record. Formally, let $X=(D_1,\ldots,D_n)\in\mathcal{D}^n$ be a tuple representing a database with $n$ entries, where $D_i\in\mathcal{D}$ is the $i$-th entry, and $\cD$ is a finite set. For $x,x'\in\mathcal{D}^n$, we write $x\sim x'$ to denote neighboring databases. Note that here, we adopt the \emph{bounded differential privacy} model~\cite{kiferNoFreeLunch2011}, where the number of records $n$ is fixed and publicly known. Thus, if $x \sim x'$, then $x'$ is obtained by replacing an entry in $x$ with a different value.

\begin{definition}[$\varepsilon$-DP~\cite{DworkCalibratingNoiseSensitivity}]
\label{def:dp}
    Given $\varepsilon> 0$, the mechanism $P_{Y\mid X}$ satisfies $\varepsilon$-DP if
    \[
    \sup_{y\in \cY}\; \max_{\substack{x,x'\in\mathcal{D}^n:\\x\sim x'}}\log \; \frac{f_{Y \mid X=x}(y)}{f_{Y \mid X=x'}(y)} \leq \varepsilon.
    \]
\end{definition}

Note that, unlike PML, DP is a property of the mechanism $P_{Y\mid X}$ alone, and its definition does not depend on the prior distribution of the data $P_X$.

Next, we define the Laplace mechanism, which is the most commonly used $\varepsilon$-DP mechanism. Let $\operatorname{Lap}(b)$ denote the Laplace distribution with mean $0$ and scale parameter $b>0$, which has variance $2b^2$. The Laplace mechanism calibrates the amount of noise based on the \emph{sensitivity} of the query.

\begin{definition}[Laplace mechanism~\cite{DworkCalibratingNoiseSensitivity}]
\label{def:laplace_mech}
Let $q : \mathcal{D}^n \to \mathbb{R}^m$ be a query with $\ell_1$-sensitivity 
\begin{equation*}
    \Delta_1(q) := \max_{x, x' \in \mathcal{D}^n : x \sim x'} {\|q(x) - q(x')\|}_1.
\end{equation*}
Suppose the elements of $N = [N_1, \ldots, N_m]^\top$ are drawn i.i.d from $\operatorname{Lap}\left(b\right)$ with $b>0$. Then, the Laplace mechanism 
\begin{equation*}
    Y = q(x) + N, \quad x \in \mathcal{D}^n, 
\end{equation*}
satisfies $\frac{\Delta_1(q)}{b}$-DP, where ${Y} = [Y_1, \ldots, Y_m]^\top$ is the output of the mechanism.
\end{definition}

In our previous works, we established two main results connecting DP and PML. First, we showed that DP is equivalent to restricting the PML of all records in all databases with independent entries~\cite{SaeidianRethinking}. Formally, let $\cQ$ denote the set of all product distributions with full support on $\cX = \cD^n$, i.e., $\mathcal Q \coloneqq \{P_{X} \colon P_{X} = \prod_{i=1}^n P_{D_i} \}$, where $P_{D_i}$ is the marginal distribution of $D_i$.

\begin{theorem}[DP as a PML Constraint {\cite{SaeidianRethinking}}]
\label{thm:dp_pml}
Given $\varepsilon > 0$, a privacy mechanism $P_{Y \mid X}$ satisfies $\varepsilon$-DP if and only if 
\begin{equation*}
    \sup_{P_X \in \mathcal Q} \; \sup_{y \in \mathcal Y} \; \max_{i \in [n]} \; \ell(D_i \to y) \leq \varepsilon. 
\end{equation*}
\end{theorem}

Second, we showed that a mechanism satisfying DP on a correlated dataset can be trivially non-private in the sense of PML. More precisely, its PML can be as large as that of a mechanism that directly releases an entry from the database without any randomization~\cite{apf24}. 

\subsection{Linear Queries}
\label{subsec:LQ}
A \emph{linear query} computes linear combinations of the counts of the input dataset. Many common queries fall into this class, such as counting queries~\cite{kamalaruban2019attributescreatedequaldmathcalxprivate}, histograms~\cite{Saeidian2025Histogram}, range queries~\cite{Li2012AnAdaptiveMechanism}, and contingency tables~\cite{Barak2007Contingency,Fienberg2010Contingency}.

Suppose \(|\mathcal{D}| = k\) and let $\{h_j\}_{j=1}^{k}$ be a collection of indicator functions, where $h_j:\mathcal{D}\to\{0,1\}$ determines whether or not a record belongs to class $j\in [k]$. 
Let $x=(d_1,...,d_n)$ be a realization of $X$. 
With a slight abuse of notation, we also use $x$ to represent the histogram of the dataset, i.e., \(x=[{x}_1,...,{x}_k]^\top \in \mathbb{N}^k\), where \({x}_j = \sum_{i=1}^n h_j(d_i)\) denotes the number of records from database $x$ belonging to class $j$.

Given a vector ${w}\in\mathbb{R}^{k}$, a single linear query calculated on $x$ can be expressed as $q(x) = {w}^\top x.$ Given a collection of $m$ linear queries $\{{w}_l\}_{l=1}^m$, the operation can be represented by the matrix product $q(x) = {W} x$, where
${W}=[{w}_1^\top,...,{w}_m^\top]^\top\in\mathbb{R}^{m\times k}$ is called the \emph{query workload}. 
The $\ell_1$-sensitivity of a query with workload $W$ is 
\begin{align*}
    \max_{x\sim x'}\| q(x) - q(x') \|_1 &= \max_{x\sim x'}\left\| {W}(x-x') \right\|_1 \\
    &= \max_{j_1,j_2\in[k]} \left\| {w}_{:,j_1} - {w}_{:,j_2} \right\|_1,
\end{align*}
where ${w}_{:,j}$ is the $j$-th column of matrix ${W}$. Thus, answering the query $q(x) = {W} x$ using the Laplace mechanism with scale $b>0$ satisfies DP  with the privacy parameter
\begin{equation}
\label{eq:dp_bound}
    \varepsilon_{\text{DP}}=\frac{\max_{j_1,j_2\in[k]} \left\| {w}_{:,j_1} - {w}_{:,j_2} \right\|_1}{b}.
\end{equation}

\section{Linear Query Release under Pointwise Maximal Leakage}
In this section, we examine the amount of information leaked when releasing linear queries using the Laplace mechanism within the PML framework. We show that, if the prior distribution of the database is suitably restricted, then the worst-case amount of information leaked is smaller than the value obtained from the standard DP analysis. This, in turn, suggests that when knowledge about the prior distribution is available, strong privacy guarantees can be achieved by adding less noise than required by DP.

\subsection{Theoretical Analysis}
Let $P_X = P_{D_1,\ldots,D_n}$ be the distribution according to which database $X$ is sampled from $\cD^n$. Recall that $\mathcal{Q}$ is the set of all product distributions over $\mathcal{D}^n$. We assume that the probability of a record falling into each class is bounded away from $0$. Specifically, let $\alpha \in (0, 1/k]$, and assume $P_X\in\mathcal{Q}_\alpha$, where
\begin{multline*}
\label{eq:prior_set}
    \mathcal{Q}_\alpha := \left\{ P_X \in \mathcal{Q} : P_{D_i}(\{d \in \mathcal{D} : h_j(d) = 1\}) \geq \alpha,\right. \\
    \left. \text{ for all } i \in [n], j \in [k] \right\}.
\end{multline*}
Note that the distributions in $\cQ_\alpha$ also satisfy $P_{D_i}(\{d \in \mathcal{D} : h_j(d) = 1\})\leq1-(k-1)\alpha$ for all $i \in [n]$ and $j \in [k]$ .
Larger values of $\alpha$ imply stronger assumptions about the data by further restricting the class of priors.

Consider a collection of linear queries answered using the Laplace mechanism, as described in Section~\ref{subsec:LQ}. Below, we characterize the PML of this mechanism under the assumption that the database is drawn according to a distribution in $\cQ_\alpha$. 
Importantly, by Theorem~\ref{thm:dp_pml}, standard DP is equivalent to bounding PML across the entire set of product distributions $\cQ$. Restricting the analysis to $\cQ_\alpha$ allows us to relax the DP setup by excluding highly skewed distributions.

\begin{theorem}
\label{thm:linear_queries_pml}
Suppose $X \sim P_X\in\mathcal{Q}_\alpha$. Let $Y=[{Y}_1,...,{Y}_m]^\top$ be the answer to a query workload released by the Laplace mechanism with scale parameter $b>0$, i.e.,
\[
    Y = W X+ N,
\]
where $N = [{N}_1, \dots, {N}_m]^\top$, $N_l \sim \operatorname{Lap}(b)$ independently for all $l\in m$, $W \in \bR^{m \times k}$ is the workload, and $X$ is the histogram of the dataset. Then, for all $i\in[n]$, distributions $P_X\in\mathcal{Q}_\alpha$ and ${y}\in\mathbb{R}^m$, the amount of information leaked about $D_i$ is upper bounded by
\begin{equation}
\label{eq:pmlbound}
    \ell(D_i\!\to\! {y})\leq \max_{\cI \subseteq[m]}\log\frac{e^{-c_{j_*}^{\cI}/b}}{\alpha\sum\limits_{j=1}^k e^{-c_j^{\cI}/b}\!+\!(1-k\alpha)e^{-c_{j^*}^{\cI}/b}},
\end{equation}
where 
\begin{equation*}
    c_j^{\cI} = \sum_{l \in \cI} {w}_{lj} - \sum_{l'\in[m]\setminus \cI} {w}_{l' j}, 
\end{equation*}
$\cI$ is a subset of $[m]$, and $j^*, j_* \in [k]$ satisfy 
\begin{gather*}
    c^\cI_{j^*} = \max_{j \in [k]} c^\cI_j, \quad c^\cI_{j_*} = \min_{j \in [k]} c^\cI_j.
\end{gather*}
\end{theorem}
\begin{IEEEproof}
Without loss of generality, we examine the information leaked about $D_1$.
Let $d_1$ be a realization of $D_1$ and suppose it belongs to class $r \in [k]$.
Let $X^-:=[\sum_{i=2}^nh_1(D_i),...,\sum_{i=2}^nh_k(D_i)]^T$ be the  histogram without the first entry.
Furthermore, let 
\[
    p_j:=P_{D_1}(\{d\in\mathcal{D}:h_j(d)=1\}),
\]
for $j\in[k]$ denote the probability that $D_1$ belongs in class $j$.
Our objective is to determine the worst-case PML across all outcomes and all prior distributions $P_X \in Q_\alpha$. Specifically, we seek the value $\sup_{P_X \in Q_\alpha} \sup_{y} \ell(D_1 \to y)$.
We fix a prior distribution $P_X$ and an outcome $y=[y_1,...,y_m]^\top\in\mathbb{R}^m$.
Using the definition of Laplace mechanism, consider the exponential of the PML expressed as
\begin{multline}
    \frac{f_{Y\mid D_1=d_1}(y)}{f_{Y}(y)}\\
    = \frac{\mathbb{E}_{X^-}\prod_{l=1}^m\exp\left(\!-\!\frac{1}{b}|y_l\!-\!w_{l,:}X^-\!-\!\sum_{j=1}^kw_{lj}h_j(d_1)|\right)}{\mathbb{E}_{X}\prod_{l=1}^m \exp\left(-\frac{1}{b}|y_l-w_{l,:}X|\right)}\\
    =\frac{\mathbb{E}_{X^-}\prod_{l=1}^m\exp\left(\!-\!\frac{1}{b}|y_l\!-\!w_{l,:}X^-\!-\!w_{lr}|\right)}{\sum_{j=1}^k p_j\mathbb{E}_{X^-}\prod_{l=1}^m \exp\left(-\frac{1}{b}|y_l-w_{l,:}X^--w_{lj}|\right)},
    \label{eq:triangle}
\end{multline}
where $w_{l,:}$ is the $l$-th row of workload $W$.
Let \( w_{l \min} = \min_j w_{lj} \) and \( w_{l \max} = \max_j w_{lj} \).
Fix a subset \( \cI \subseteq [m] \),  and suppose \( y \) satisfies \( y_l \leq nw_{l \min} \) for \( l \in \cI \) and \( y_l \geq nw_{l \max} \) for \( l \in [m]\setminus \cI \). 
As argued in Lemma~\ref{lma:1} in Appendix~\ref{sec:ProofofTheorem}, we can remove the absolute values and analyze the expression only in regions $y_l\in(-\infty,nw_{l\min}]$ and $y_l\in[nw_{l\max},\infty)$ for $l\in [m]$.
The RHS of~\eqref{eq:triangle} is upper bounded by
\begin{align*}
    &\sup_{\cI\subseteq[m]}\frac{\mathbb{E}_{X^-}\Bigl[\prod_{l\in \cI}\exp\left(\frac{1}{b}(y_l\!-\!w_{l,:}X^-\!-\!w_{lr})\right)}{\sum_{j=1}^k p_j\Bigl\{
    \mathbb{E}_{X^-}\Bigl[\prod_{l\in \cI} \exp\left(\frac{1}{b}(y_l-w_{l,:}X^--w_{lj})\right)}\\
    &\quad\quad\quad\quad\quad\frac{\prod_{l'\in[m]\setminus \cI}\exp\left(-\frac{1}{b}(y_{l'}\!-\!w_{l',:}X^-\!-\!w_{l'r})\right)\Bigr]}{\prod_{l'\in[m]\setminus \cI} \exp\left(-\frac{1}{b}(y_{l'}-w_{l',:}X^--w_{l'j})\right)\Bigr]\Bigr\}}\\
    =&\sup_{\cI\subseteq[m]}\;
    \frac{
    \mathbb{E}_{X^-}\!\Bigl[
    \prod_{l\in \cI}
    e^{y_l/b}\,e^{-w_{l,:}X^-/b}\,e^{-w_{lr}/b}
    \Bigr.
    }{
    \sum_{j=1}^k p_j\Bigl\{
    \mathbb{E}_{X^-}\!\Bigl[
    \prod_{l\in \cI}
    e^{y_l/b}\,e^{-w_{l,:}X^{-}/b}\,e^{-w_{lj}/b}
    \Bigr.
    \Bigr.
    }
    \\
    &
    \qquad\qquad\qquad\qquad\frac{
    \prod_{l'\in [m]\setminus \cI}
    e^{-y_{l'}/b}\,e^{w_{l',:}X^-/b}\,e^{w_{l'r}/b}
    \Bigr]
    }{
    \prod_{l'\in [m]\setminus \cI}
    e^{-y_{l'}/b}\,e^{w_{l',:}X^-/b}\,e^{w_{l'j}/b}
    \Bigr]
    \Bigr\}
    }\\
    =&\sup_{\cI\subseteq[m]}\;
    \frac{
    \exp\!\left(-\frac{1}{b}\Bigl(\sum_{l\in \cI} w_{lr}-\sum_{l'\in [m]\setminus \cI} w_{l'r}\Bigr)\right)
    }{
    \sum_{j=1}^k p_j\,\exp\!\left(-\frac{1}{b}\bigl(\sum_{l\in \cI} w_{lj}-\sum_{l'\in [m]\setminus \cI} w_{l'j}\bigr)\right)
    }
\end{align*}
Let \( c_j^{\cI} := \sum_{l \in \cI} w_{lj} - \sum_{l'\in[m]\setminus \cI} w_{l' j} \). So far, we have shown that
\begin{equation*}
    \frac{f_{Y\mid D_1=d_1}(y)}{f_{Y}(y)}\leq\sup_{\cI\subseteq[m]}\frac{\exp \left( -\frac{1}{b}c_{r}^{\cI} \right)}
    {\sum_{j=1}^kp_j\exp(-\frac{1}{b}c_j^\cI)}
    \label{eq:simple}
\end{equation*}

Finally, with Lemma~\ref{lma:2} in Appendix~\ref{sec:ProofofTheorem}, observing that the above argument holds for $d_1$ belonging to any class, we get
\[
\exp(\ell(D_i\!\to\! y))\leq \sup_{\cI\subseteq[m]} \frac{e^{-c_{j_*}^{\cI}/b}}{\alpha\sum_{j=1}^k e^{-c_j^{\cI}/b}+(1-k\alpha)\,e^{-c_{j^*}^{\cI}/b}},
\]
where \(j_*\) is such that \( c_{j_*}^{L} = \min_j c_j^{L} \). 
\end{IEEEproof}

The bound in Theorem~\ref{thm:linear_queries_pml} is tight, so there exists $i \in [n]$, $y \in \bR^m$, and $P_X \in \cQ_\alpha$ such that $\ell(D_i \to y)$ equals the right-hand side. However, \eqref{eq:pmlbound} can be computationally expensive to evaluate, especially for large workloads, as it involves calculating an expression for all subsets $\cI \subseteq [m]$. For this reason, we further upper bound \eqref{eq:pmlbound} to obtain an expression that is easier to compute.

\begin{figure*}[ht!]
  \centering
  \subfigure[Histogram Query]{
    \includegraphics[width=0.31\textwidth]{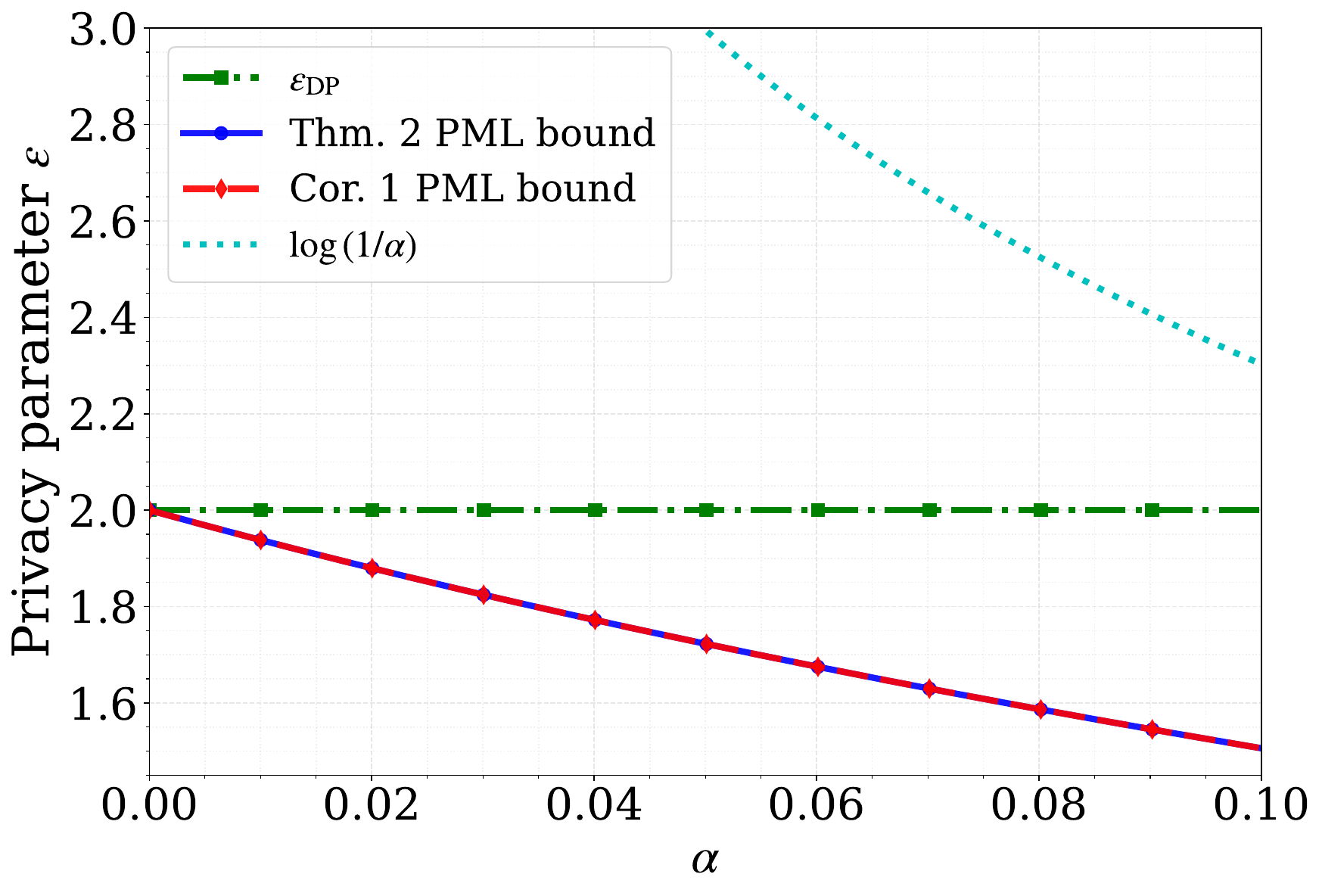}
    \label{fig:res_hist}
  }
  \hfil
  \subfigure[Range Query]{
    \includegraphics[width=0.31\textwidth]{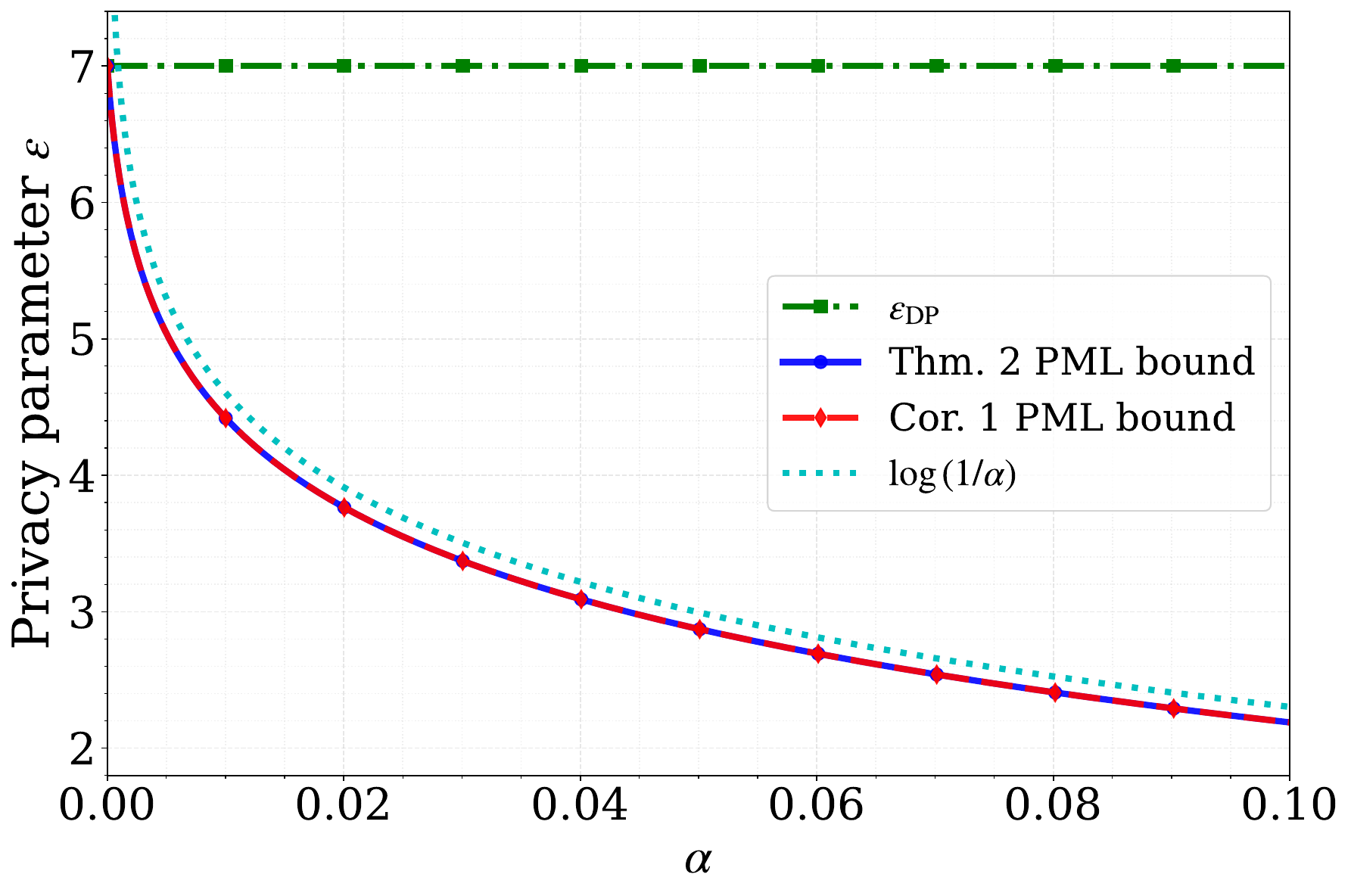}
    \label{fig:res_range}
  }
  \hfil
  \subfigure[Difference Query]{
    \includegraphics[width=0.31\textwidth]{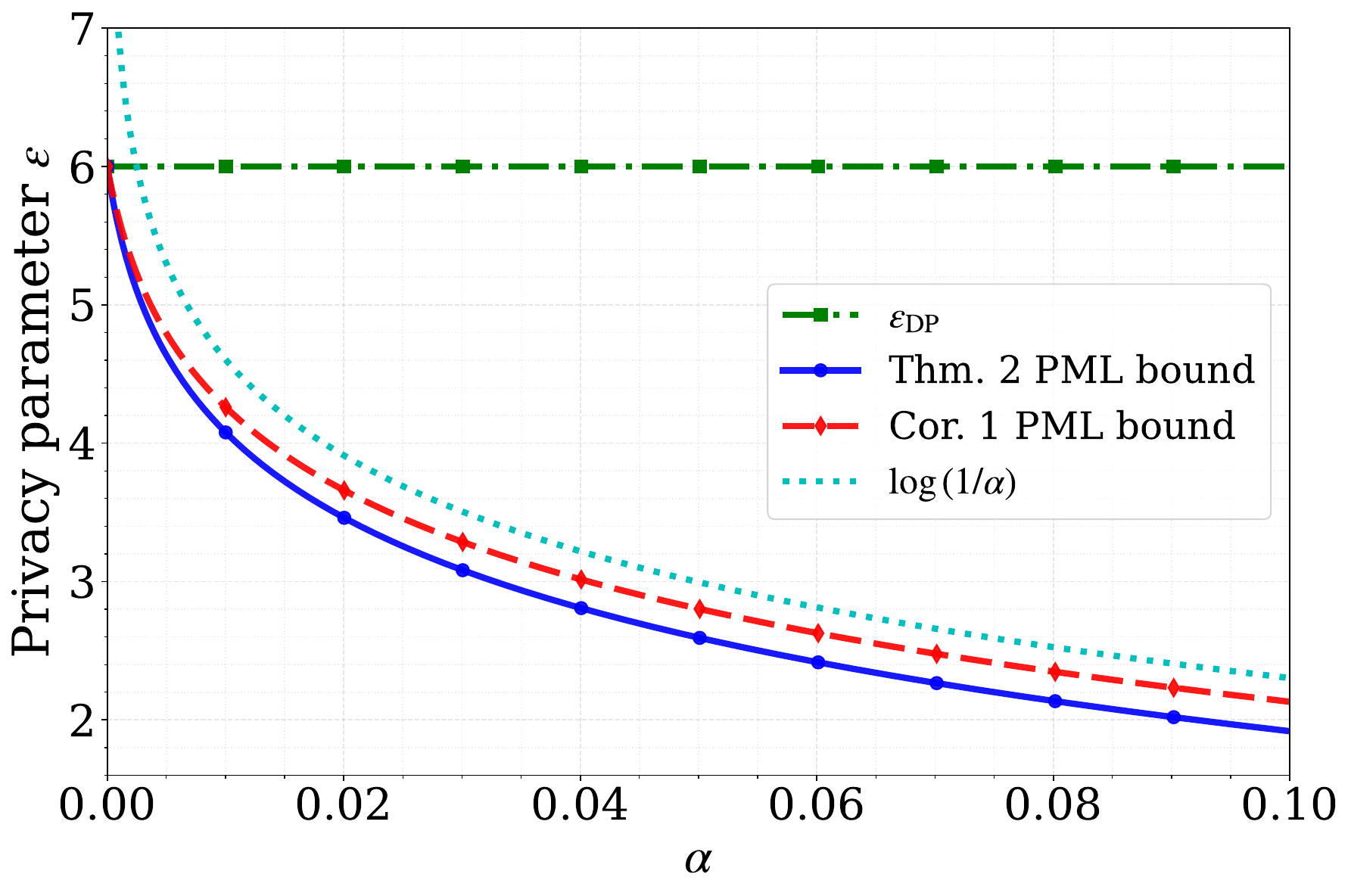}
    \label{fig:res_haar}
  }
  
  \caption{Comparison of privacy bounds for different workload types (leakage vs. prior parameter $\alpha$).  
  }
  \label{fig:leakage_results}
\end{figure*}

\begin{figure}
  \centering
  \includegraphics[width=0.45\textwidth]{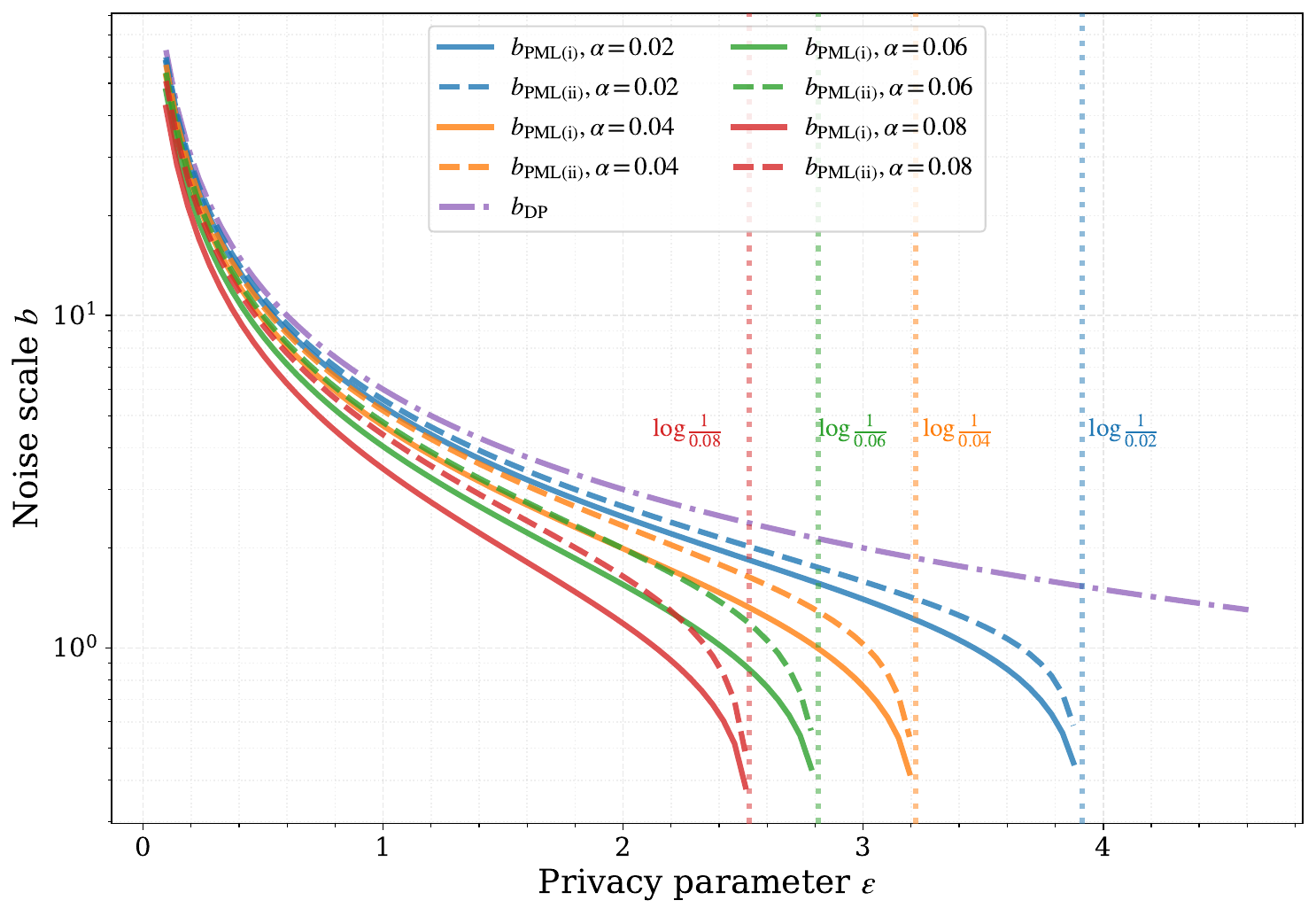}
  
  \caption{Noise scale $b$ vs. privacy parameter $\varepsilon$.
  This figure illustrates the minimum noise scale $b$ required to ensure a target privacy leakage $\varepsilon$ for the difference query workload~\eqref{eq:haar_kronecker}.
  The solid ($b_{\text{PML(i)}}$), dashed ($b_{\text{PML(ii)}}$), and dash-dotted ($b_{\text{DP}}$) lines denote the minimum noise scales required for Thm.~\ref{thm:linear_queries_pml}, Cor.~\ref{crl:simplebound}, and $\varepsilon_{\text{DP}}$ to be at most $\varepsilon$, respectively.
  }
  \label{fig:minb}
\end{figure}

\begin{corollary}
\label{crl:simplebound}
Consider the setup of Theorem~\ref{thm:linear_queries_pml}. Then, we have 
\begin{equation}
\label{eq:simple_bound}
    \ell(D_i \!\to \!y) \!\leq \! \max_{j_1,j_2\in[k]}\log \bigg(\!\alpha\sum\limits_{j=1}^k e^{-\Delta_{j,j_1}} \! + (1-k\alpha)e^{-\Delta_{j_1,j_2}} \! \bigg)^{-1},
\end{equation}
where $\Delta_{j,j'} \coloneqq\frac{\|{w}_{:,j}-{w}_{:,j'}\|_1}{b}$ for $j,j'\in[k]$. Furthermore, the bound is tight if and only if there exists a subset $\cI^*\subseteq[m]$ in such that for all $j,j'\in[k], {w}_{l,j'}-{w}_{l,j'}\leq0$ for all $l\in \cI^*$ and ${w}_{l,j}-{w}_{l,j'}\geq0$ for all $l\in[m]\setminus \cI^*$.
\end{corollary}
\begin{IEEEproof}
    See Appendix~\ref{sec:ProofofSimplified}. 
\end{IEEEproof}

Next, let us discuss how our bounds in Theorem~\ref{thm:linear_queries_pml} and Corollary~\ref{crl:simplebound} behave as a function of $\alpha$ and how they compare to $\varepsilon_{\text{DP}}$ in \eqref{eq:dp_bound}. 
\begin{remark}
\label{rmk:cmprDP}
The PML bounds of Theorem~\ref{thm:linear_queries_pml} and Corollary~\ref{crl:simplebound} have the following properties: 
\begin{enumerate}[label=(\roman*)]
    \item It is straightforward to verify that the right-hand sides of \eqref{eq:pmlbound} and \eqref{eq:simple_bound} are decreasing in $\alpha$. This behavior is a recurring theme in PML-based privacy analysis: Larger values of $\alpha$ exclude the more skewed distributions from protection, and privacy is generally easier to guarantee when the data is more uniformly distributed. 

    \item We can further upper bound \eqref{eq:simple_bound} as follows:
    \begin{align*}
         &\ell(D_i \to y)  \\
         &\leq \!\max\limits_{j_1,j_2\in[k]}\log\!\bigg(\!\alpha + \!\alpha\!\sum\limits_{j \neq j_1} \!e^{-\Delta_{j,j_1}} \! + \! (1-k\alpha)e^{-\Delta_{j_1,j_2}}\!\bigg)^{-1}\\
         &\leq \log \frac{1}{\alpha}. 
    \end{align*}
    Note that this is simply the trivial PML upper bound in~\eqref{eq:trivial} for $P_X \in \cQ_\alpha$. This bound holds regardless of the privacy mechanism used, including the case where the query is answered without any randomness. Thus, for $P_X \in \cQ_\alpha$, any DP guarantee with $\varepsilon_{\text{DP}} \geq \log (1/\alpha)$ is vacuous (in the sense of PML). 

    \item As $\alpha \to 0$, both the RHS of \eqref{eq:pmlbound} and \eqref{eq:simple_bound} approach 
    \begin{equation}
    \label{eq:limit_alpha_zero}
         \max_{\cI \in [m]} \Big(\frac{c_{j^*} - c_{j_*}}{b} \Big) = \max_{j,j'} \Delta_{j,j'} = \varepsilon_{\text{DP}}.  
    \end{equation}
    Furthermore, both bounds are strictly smaller than $\varepsilon_{\text{DP}}$ for all $\alpha \in (0, 1/k]$. Observe that \eqref{eq:limit_alpha_zero} is consistent with, and was expected from Theorem~\ref{thm:dp_pml}. 
    \label{property:compareDP}
    \begin{IEEEproof}
        See Appendix~\ref{sec:ProofofProperty}.
    \end{IEEEproof}
\end{enumerate}
\end{remark}

\subsection{Numerical Evaluations}
Here, we empirically evaluate our bounds across four representative classes of linear queries. We construct workload matrices $W \in \mathbb{R}^{8 \times 8}$ for the following scenarios:
\begin{enumerate}[label=(\roman*)]
    \item A \textit{histogram query}, where $W=I_8$ is the identity matrix.
    \item A collection of \textit{range queries}.
    Each row $l$ corresponds to a range query defined by an interval $[L_l, R_l]$, where $1 \leq L_l \leq R_l \leq k$. 
    Specifically, the entry $w_{lj} = 1$ if $j \in [L_l, R_l]$ and $w_{lj} = 0$ otherwise. 
    The interval endpoints $L_l$ and $R_l$ for each query are sampled uniformly at random: $L_l$ is drawn uniformly from $\{0, 1, \ldots, k-1\}$, and $R_l$ is drawn uniformly from $\{L_l, L_l+1, \ldots, k-1\}$, ensuring that each row contains exactly one contiguous block of ones.
    
    \item A collection of \textit{difference queries} constructed using the unnormalized Haar wavelet transformation matrix. 
    $W$ is structured hierarchically to capture information at multiple resolutions, defined as
    {
    \begin{align}
    W &=
    \begin{bmatrix}
    \mathbf{1}_8 \\[-0.5pt]
    h \otimes \mathbf{1}_4 \\[-0.5pt]
    I_2 \otimes h \otimes \mathbf{1}_2 \\[-0.5pt]
    I_4 \otimes h
    \end{bmatrix}, 
    \text{with } h=[1, -1],
    \label{eq:haar_kronecker}
    \end{align}}
    where $\mathbf{1}_p$ denotes the $p$-dimensional all-ones vector, and $\otimes$ denotes the Kronecker product.
    \label{workload:diff}
\end{enumerate}
All queries are released via the Laplace mechanism with the noise scale $b=1.0$.

Figure~\ref{fig:leakage_results} illustrates the privacy leakage as a function of the prior parameter $\alpha$.
As expected, for all query types, the Thm.~\ref{thm:linear_queries_pml} PML bound is strictly tighter than the context-free DP budget ($\varepsilon_{\text{DP}}$) whenever prior knowledge is present ($\alpha > 0$).
Furthermore, the Cor.~\ref{crl:simplebound} PML bound, while slightly looser than the exact PML calculation, effectively tracks the leakage trend and remains tighter than the DP baseline.
Notably, the gap between the PML and DP bounds is most pronounced in the high-entropy regime (large $\alpha$), indicating that DP overestimates risk when the prior is close to uniform.

Figure~\ref{fig:minb} presents the same trade-off from a utility perspective.
Instead of fixing the noise scale $b$, we plot the minimum noise $b$ required to satisfy a target privacy parameter $\varepsilon$.
Consistent with the leakage analysis, guaranteeing a specific $\varepsilon$-PML requires less noise than satisfying $\varepsilon$-DP.
Crucially, as the privacy budget approaches the intrinsic uncertainty of the prior (i.e., $\varepsilon \to \log(1/\alpha)$), the required noise vanishes to zero, a regime that pure DP cannot capture.

\section{Conclusions and Future Work}
\label{sec:conclusion}

In this paper, we revisited the privacy analysis of the Laplace mechanism for linear query workloads through the lens of pointwise maximal leakage (PML). 
By accounting for the minimum probability of data classes, we derived a tight leakage bound that is strictly stronger than context-free DP analysis and naturally converges to the DP budget as this prior knowledge vanishes. 

Building on this analytical foundation, our future work aims to pivot from analysis to synthesis by developing novel privacy mechanisms explicitly designed under PML constraints. 
A promising direction is to integrate the \emph{matrix mechanism}~\cite{li2015matrix} framework with PML-based optimization. 
While standard matrix mechanisms optimize the query strategy under differential privacy constraints, reformulating this optimization problem with PML constraints could yield significant utility gains. 
We plan to theoretically characterize the optimal mechanism structures for linear queries under PML and empirically validate their improvements against state-of-the-art DP baselines.

\clearpage
\printbibliography

\clearpage
\appendices

\section{Auxiliary Lemmas for Proof of Theorem~\ref{thm:linear_queries_pml}}
\label{sec:ProofofTheorem}
\begin{lemma}
    In order to find the worst-case PML, it is sufficient to analyze \eqref{eq:triangle} only in regions $y_l\in(-\infty,nw_{l\min}]$ and $y_l\in[nw_{l\max},\infty)$ for $l\in [m]$.
    \label{lma:1}
\end{lemma}
\begin{IEEEproof}
    Note in the numerator, we have
    \begin{multline*}
    \exp\Bigl(-\tfrac1b \,\lvert y_l-{w}_{l,:}{X}^- - w_{lr}\rvert\Bigr)
    \\\le \!\min\!\Bigl\{\exp\!\bigl(-\tfrac1b(y_l\!-\!{w}_{l,:}{X}^-\!-\!w_{lr})\bigr),\,
    \exp\!\bigl(\tfrac1b(y_l\!-\!{w}_{l,:}{X}^- \!- \!w_{lr})\bigr)\Bigr\},
    \end{multline*}
    for all $l\in[m]$. 
    Furthermore, since we have \( 0 \leq \sum_{i=1}^{n} h_j(D_i) \leq n \text{ for all }j\) and \( \sum_{j=1}^{k} \sum_{i=1}^{n} h_j(D_i) = n \), the $l$-th outcome before adding noise $w_{l,:}X=\sum_{j=1}^{N} w_{lj} \sum_{i=1}^{n} h_j(D_i)$ satisfies
    \begin{align}
    nw_{l \min} \leq \sum_{j=1}^{k} w_{lj} \sum_{i=1}^{n} h_j(D_i) \leq nw_{l \max}.
    \label{eq:propor}
    \end{align}
    Note that in the denominator, the mapping $y_l \mapsto -\frac{1}{b} \left| y_l - w_{l,:}X \right|$ is increasing on $(-\infty,nw_{l\min}]$ and decreasing on $[nw_{l\max},\infty)$. 
    Hence, these regions maximize the ratio by minimizing the denominator.
\end{IEEEproof}

\begin{lemma}
    For \( p_j \geq \alpha \), \( \sum_j p_j = 1 \), the minimum of $\sum_{j=1}^k p_j e^{-c_j^{\cI}/b}$ is
    \[
        \alpha\sum_{j=1}^k e^{-c_j^{\cI}/b}+(1-k\alpha)\,e^{-c_{j^*}^{\cI}/b},
    \]
    where \(j^*\) satisfies \( c_{j^*}^{\cI} = \max_j c_j^{\cI} \).
    \label{lma:2}
\end{lemma}
\begin{IEEEproof}
    Using \(p_j=\alpha+(p_j-\alpha)\) and \(\sum_{j=1}^k p_j=1\), we have
    \[
    \sum_{j=1}^k p_j e^{-c_j^{\cI}/b}
    = \alpha\sum_{j=1}^k e^{-c_j^{\cI}/b}
      + \sum_{j=1}^k (p_j-\alpha)\,e^{-c_j^{\cI}/b},
    \]
    with \(p_j-\alpha\ge 0\) and \(\sum_{j=1}^k (p_j-\alpha)=1-k\alpha\).
    Since \(j^*=\arg\max_j c_j^{\cI}\), for all \(j\) we have \(e^{-c_j^{\cI}/b}\ge e^{-c_{j^*}^{\cI}/b}\).
    Therefore,
    \[
    \sum_{j=1}^k (p_j-\alpha)\,e^{-c_j^{\cI}/b}
    \;\ge\;
    (1-k\alpha)\,e^{-c_{j^*}^{\cI}/b},
    \]
    which yields
    \[
    \sum_{j=1}^k p_j e^{-c_j^{\cI}/b}
    \;\ge\;
    \alpha\sum_{j=1}^k e^{-c_j^{\cI}/b}+(1-k\alpha)\,e^{-c_{j^*}^{\cI}/b}.
    \]
\end{IEEEproof}

\section{Proof of Corollary~\ref{crl:simplebound}}
\label{sec:ProofofSimplified}
In~\eqref{eq:triangle}, we apply the triangle inequality $|y_l-w_{l,:}X^--w_{lj}|\leq|y_l-w_{l,:}X^--w_{lr}|+|w_{lj}-w_{lr}|$ to the denominator, and we get
\begin{align*}
   \frac{f_{Y\mid D_1=d_1}(y)}{f_{Y}(y)}&\le
\frac{
  \mathbb{E}_{X^-}\prod_{l=1}^m
  \exp\!\left(-\frac{1}{b}\bigl|y_l-w_{l,:}X^--w_{lr}\bigr|\right)
}{
  \begin{aligned}[t]
  &\sum_{j=1}^kp_j\!\left(\prod_{l=1}^m \exp\!\left(-\frac{1}{b}\bigl|w_{lj}-w_{lr}\bigr|\right)\right)\\
  &\times \mathbb{E}_{X^-}\prod_{l=1}^m
    \exp\!\left(-\frac{1}{b}\bigl|y_l-w_{l,:}X^--w_{lr}\bigr|\right)
  \end{aligned}
}\\
&=\Bigl(\sum_{j=1}^k
p_j\,\exp\!\bigl(-1/b\|w_{:,j}-w_{:,r}\|\bigr)\Bigr)^{-1}.
\end{align*}
Then, following the same reasoning as in the proof of Lemma~\ref{lma:2}, the expression is at most
\[
    \max_{j_1,j_2\in[k]}\Bigl\{\alpha\sum_{j=1}^k e^{-\frac{\|w_{:,j}-w_{:,j_1}\|_1}{b}}+(1-k\alpha)\,e^{-\frac{\|w_{:,j_2}-w_{:,j_1}\|_1}{b}}\Bigr\}^{-1},
\]
and we obtain the desired expression.

\section{Proof of Property~\ref{property:compareDP}}
\label{sec:ProofofProperty}
To show the bound in Theorem~\ref{thm:linear_queries_pml} is tighter than $\varepsilon_{\text{DP}}$, rewrite the exponential of the bound as 
\[\max_{\cI \subseteq[m]}\Bigl\{\alpha \sum_{j=1}^k\exp\Bigl(\frac{c_{j'}^{\cI}-c_j^{\cI}}{b}\Bigr)+(1-k\alpha)\exp\Bigl(\frac{c_{j'}^{\cI}-c_{j^*}^{\cI}}{b}\Bigr)\Bigr\}^{-1}.\]
Because for any \( \cI \), $c_{j^*}^{\cI}-c_{j'}^{\cI}\geq c_{j}^{\cI}-c_{j'}^{\cI}$ for all $j$, the above expression is at most \( \exp \Bigl( \frac{c_{j^*}^{{\cI}} - c_{j'}^{{\cI}}}{b} \Bigr) \) which is obtained in the limit when $\alpha\to 0 $.
Recall that \(c_j^{\cI} = \sum_{l \in {\cI}} w_{lj} - \sum_{l' \in[m]\setminus {\cI}} w_{l'j}\), and compute
\begin{align*}
    &c_{j^*}^{\cI} - c_{j'}^{\cI} \\
    =& \Bigl( \sum_{l \in {\cI}} w_{l j^*} - \sum_{l' \in[m]\setminus {\cI}} w_{l' j^*} \Bigr) - \Bigl( \sum_{l \in {\cI}} w_{l j'} - \sum_{l' \in[m]\setminus {\cI}} w_{l' j'} \Bigr) \\
    =& \sum_{l \in {\cI}} (w_{l j^*} - w_{l j'}) + \sum_{l' \in[m]\setminus {\cI}} (w_{l' j'} - w_{l' j^*}).
\end{align*}
Applying the triangle inequality, we get
\begin{align*}
    \bigl| c_{j^*}^{\cI} - c_{j'}^{\cI} \bigr| 
    &= \Bigl| \sum_{l \in {\cI}} (w_{l j^*} - w_{l j'}) + \sum_{l' \in[m]\setminus {\cI}} (w_{l' j'} - w_{l' j^*}) \Bigr|\\ 
    &\leq \sum_{l \in {\cI}} \left| w_{l j^*} - w_{l j'} \right| + \sum_{l' \in[m]\setminus {\cI}} \left| w_{l' j^*} - w_{l' j'} \right|\\ 
    &= \sum_{l=1}^m \left| w_{l j^*} - w_{l j'} \right| = \|w_{:,j^*} - w_{:,j'}\|_1\\ 
    &\leq \max_{j_1, j_2} \|w_{:,j_1} - w_{:,j_2}\|_1.
\end{align*}
Thus:
\begin{equation}
     c_{j^*}^{\cI} - c_{j'}^{\cI}  \leq \max_{j_1, j_2} \|w_{:,j_1} - w_{:,j_2}\|_1.
    \label{eq:leql1}
\end{equation}
Take two columns \(j_1, j_2 \in [k]\) such that \(\|w_{:,j_1} - w_{:,j_2}\|_1 = \max_{j', j''} \|w_{:,j'} - w_{:,j''}\|_1\). Let \({\cI} := \{ l \in [m] : w_{l j_1} \geq w_{l j_2} \}\). 
Then,
\begin{align}
    c_{j_1}^{\cI} - c_{j_2}^{\cI} &= \sum_{l \in {\cI}} (w_{l j_1} - w_{l j_2}) - \sum_{l' \in[m]\setminus {\cI}} (w_{l' j_1} - w_{l' j_2})\nonumber\\ 
    &= \sum_{l: w_{l j_1} > w_{l j_2}} |w_{l j_1} - w_{l j_2}| + \sum_{l: w_{l j_1} < w_{l j_2}} |w_{l j_1} - w_{l j_2}|\nonumber\\ 
    &= \sum_{l=1}^m |w_{l j_1} - w_{l j_2}| = \|w_{:,j_1} - w_{:,j_2}\|_1.
    \label{eq:leql2}
\end{align}
Equation~\eqref{eq:leql2} establishes that the equality in~\eqref{eq:leql1} holds for any workload. Specifically, \(\max_{{\cI}} \left( c_{j^*}^{\cI} - c_{j'}^{\cI} \right) = \max_{j_1, j_2} \|w_{:,j_1} - w_{:,j_2}\|_1\), where the maximizing \({\cI}\) is given by \({\cI} = \{ l \in [m] : w_{l j_1} \geq w_{l j_2} \}\), and \(j_1, j_2\) are the workload columns with the largest \(\ell_1\) distance. Consequently, the bound in Theorem~\ref{thm:linear_queries_pml} reaches its maximum \( \max_{{\cI}\subseteq[m]}\exp \Bigl( \frac{c_{j^*}^{{\cI}} - c_{j'}^{{\cI}}}{b} \Bigr) \)  as \(\alpha \to 0\), which is equal to $\varepsilon_{\text{DP}}= \frac{\max_{j_1, j_2\in[k]} \| w_{:,j_1} - w_{:,j_2} \|_1}{b}$. 
For \(\alpha > 0\), however, the bound is strictly tighter than \(\varepsilon_{\text{DP}}\). 
Notably, when \(\alpha > 0\), the \({\cI}\) that maximizes~\eqref{eq:pmlbound} may differ from the one described above.




\end{document}